# Fractal and multifractal descriptors restore ergodicity broken by non-Gaussianity in time series

Damian G. Kelty-Stephen[1] and Madhur Mangalam[2]

[1]*Department of Psychology, State University of New York at New Paltz, New Paltz, NY, USA*

[2]*Department of Physical Therapy, Movement and Rehabilitation Sciences, Northeastern University, Boston, MA, USA*

**ORCIDs:**

Damian G. Kelty-Stephen (0000-0001-7332-8486)

Madhur Mangalam (0000-0001-6369-0414)

**Authors for correspondence:**

Damian G. Kelty-Stephen

e-mail: keltystd@newpaltz.edu

Madhur Mangalam

e-mail: m.manglam@northeastern.edu

**Abstract**

Ergodicity breaking is a challenge for biological and psychological sciences. Ergodicity is a necessary condition for linear causal modeling. Long-range correlations and non-Gaussianity characterizing various biological and psychological measurements break ergodicity routinely, threatening our capacity for causal modeling. Long-range correlations (e.g., in fractional Gaussian noise, a.k.a. "pink noise") break ergodicity—in raw Gaussian series, as well as in some but not all standard descriptors of variability, i.e., in coefficient of variation (*CV*) and root mean square (*RMS*) but not standard deviation (*SD*) for longer series. The present work demonstrates that progressive increases in non-Gaussianity conspire with long-range correlations to break ergodicity in *SD* for all series lengths. Meanwhile, explicitly encoding the cascade dynamics that can generate temporally correlated non-Gaussian noise offers a way to restore ergodicity to our causal models. Specifically, fractal and multifractal properties encode both scale-invariant power-law correlations and their variety, respectively, both of which features index the underlying cascade parameters. Fractal and multifractal descriptors of long-range correlated non-Gaussian processes show no ergodicity breaking and hence, provide a more stable explanation for the long-range correlated non-Gaussian form of biological and psychological processes. Fractal and multifractal descriptors offer a path to restoring ergodicity to causal modeling in these fields.

**Keywords:** causality, far-from-equilibrium, heterogeneity, lognormality, nonequilibrium, stationarity

# 1. Introduction

Ergodicity remains an exotic-sounding construct in biological and psychological sciences, limited to the footnotes or the chapters left unassigned in the statistics curricula for these scientific training. Of course, it is hard to cover everything. The statistics curriculum in these fields must necessarily compromise with the more specific content. Science students need the dry, ethereal abstraction of statistical modeling—for example, medians and means, regressions and residuals—but they need to balance that with the wet organisms—for example, blood cells and button presses, foot slips, and filaments. Indeed, we cannot do without either side in the biological and psychological sciences: we need the modeling abstractions to learn what we wish to know about cause-and-effect in the concrete, tangible facts of life and mind. Our measurements catch changes in biology and psychology's wet, analog processes, then encode those measured changes into models. We pin down a sense of causality when the resulting shape of our models fits into a statistical space within those measured changes. The cycling between measurement and model is centrally crucial to the scientific method for a great deal of its empirical progress [1]. We need the right measurements of the right systems to populate our statistical models most effectively. Unfortunately, modeling alone may not suffice: a statistical logic to exclude domain-specific content can be barren for many biological and psychological investigations [2]. So, balance the statistical curricula we must, and ergodicity has perennially fallen from the chopping block [3–9].

## 1.1. Ergodicity is necessary for cause and likely absent from raw measurements of living/thinking systems

One problem with the still-exotic status of ergodicity is that it is central to all dominant modeling of cause-and-effect. So, the omission of ergodicity may be a sacrifice that scientific training can ill afford. Ergodicity is the expectation that large samples should converge towards averages and, in a quick logical succession, that the average behavior across a group replicates so well that it represents or predicts the individual behavior. Any modern notion of "cause" as a factor with a reliable, replicable effect on average depends on—at the root—average behavior giving us a privileged view of the causal effect. Indeed, most null-hypothesis testing hinges on a Gaussian-approximating sampling distribution encoding the probability of sample means in samples of stipulated size. The complementary approach of estimating a standardized effect size depends on scaling a difference in average by a stable standard deviation, measuring how far the Gaussian distribution has moved in terms of its increments. The role of the average does not go away once we level up to explicitly "causal" statistical analysis. It only becomes sublimated into ever more refined statistical approaches (e.g., Granger causality, vector autoregression, structural equation modeling) that require a stable average. As causal modeling incorporates more measured variables and more estimated coefficients, it only invests itself more deeply into a broader set of averages—all of which averages, again, must be stable. Modeling causality requires the balance of many Gaussian distributions, each clinging to their stable average, and we expect all of these models to replicate.

Another problem with the still-exotic status of ergodicity is that biologists and psychologists are mainly ill-equipped to recognize the widespread failure of ergodicity to hold in measurements. As ergodicity slips from the pedagogical chopping block, it leaves the biologist and psychologist with a lopsided portfolio. Experience with what and how to measure far outstrips the ability to evaluate which models will fit the measured data. This problem also falls disproportionately on the biological and psychological sciences because these fields focus primarily, by design, on systems that are primarily interesting for their capacity for nonstationary, creative, context-dependent, and constraint-breaking ways.

The physical and chemical sciences that may pursue empirical research while constraining the model systems at atomic and subatomic levels can, in effect, tune the desired equilibrium into their analog processes. As soon as we begin to entertain systems whose degrees of freedom have the liberty of multiple scales of activity, we see a stark divergence between the possibility and the probability of converging around an average.

## 1.2. What is ergodicity?

The question "What is ergodicity" turns out to be timely for all scientists. Just as ergodicity is getting some of its first looks from the biological and psychological sciences [8], ergodicity has recently elicited a newer look from the physical sciences. The traditional view was that ergodicity was a convergence of two averages: first, the average of a whole sample of processes at the same point in time,

$$\langle x(t) \rangle_N = \frac{1}{N} \sum_i^N x_i(t) \quad (1)$$

for $x_i$ representing position of the $i$th of $N$ realizations $x(t)$ at time $t$, and second, the average of a single process over time,

$$\overline{x_{\Delta t}} = \frac{1}{\Delta t} \int_t^{t+\Delta t} x(s)\,ds. \quad (2)$$

That is, the traditional view of ergodicity had been that the average behavior of the sample would converge on the average trajectory of the individual. A newer interpretation raises a stricter standard for ergodicity through the criterion of "mixing". Mixing is the property of independence across time such that a stochastic process would exhibit equal probabilities of all values across all times. Averages can fail to converge even under weak mixing, allowing even stationary processes to break ergodicity [10].

A more tangible example of ergodicity and its failure is in order until readers from biological and psychological domains become more comfortable with the finer points of average convergence and mixing. To the behavioral scientist, a good approximation of ergodicity is when average performance across a sample of participants represents the performance of the individual participant. Conversely, nonergodic processes exhibit statistics that change with time, for example, when the performance of a participant diverges from the average performance of the sample. Gambling offers a classic example of nonergodicity [6]: after ever flip of a fair coin, you win 50% of your current wealth of $1 for heads, and you lose 40% for tails. Assuming a sample of players with infinite wealth and infinite time, the average winnings increase exponentially (figure 1). However, in practice, the individual gamblers lose their winnings and quickly lose the disposable resources needed to stay in the game. The average increase in winning is only the consequence of a lucky minority.

## 1.3 Ergodicity might be fundamentally incompatible with the striving and context-sensitivity of life and mind

Life is, by definition, a forward-looking, past-remembering, context-sensing, exploring, social, and often self-promoting process—it can recur and oscillate, and it can regress to the mean [11–13]. Nonetheless, what can sometimes happen is not the rule for what always happens: life carries information from the past to the present, producing strongly autocorrelated and sometimes trended or biased trajectories. The conditions for expecting regression to the mean are fragile and specific to a stationary, often Gaussian process with independence among constituent causal factors [14]. Here we see an echo of our earlier point: ergodicity is the stability of averages one can only expect when the constituent causes themselves are stable and independent. We rely on the converse of this point in our linear statistical models of cause: we use averages to speak to the independent causal factors that we assume compose our measured variation.

Life and mind—characterized by interactions amongst constituent causes and dependencies across time—ransack many of these premises. Furthermore, the mathematical and observable consequence of this fact is as clear as it can be: lognormal deviations from Gaussian distributions and correlations across time lead standard deviation to grow across all observable spatial and temporal scales [15–20]. It might easily sound like biological and psychological sciences never had any hope of finding ergodicity in all but the most inert and passive preparations of biological tissues, and it might easily sound like we would need to immobilize life on a Procrustean bed of enforced equilibrium before ever finding a trustworthy, replicable cause in these fields. Causality was only ever accessible to science for the most lifeless and mindless of behaviors.

### 1.4. Paths forward: The sources of nonergodicity in our raw measurements may themselves offer ergodic candidates for causal modeling

All is not lost, though, and a growing body of research promises to detail both the problem of nonergodicity and possible journeys towards its resolution. What is more the areas of physics and chemistry that are ready to entertain systems at, near, and far-from-equilibrium provide a ready toolset in this endeavor [8]. For instance, we can be clear-eyed and now realize that life and mind break ergodicity pervasively, from the molecular dynamics in a cell [21–25] to extracellular dynamics [26,27], from physiology [28] to neural dynamics [29,30], and from perception to cognition [31–37]. We can begin to quantify ergodicity breaking to examine both the factors likely to break ergodic. For instance, we can find that measures of postural sway breaks ergodicity even while participants attempt to stand perfectly still, and we have been able to show that stronger temporal correlations in postural sway breaks ergodicity [8].

Ergodicity breaks as our measures depart from the classical assumptions of linear statistical modeling, that is, from independent selections from a Gaussian distribution with stable mean and always identical variance ("white noise"). White noise is stationary, but more importantly, it is "mixing" which is the state of being independent of time and so homogeneous across all time lags. Without even requiring nonstationarity, we can use temporal correlations to weaken this mixing property, as in the case of fractional Gaussian noise (fGn), whose autocorrelation shows a power-law decay of autoregressive coefficient $\rho$ for lag $k$ as $\rho k = \frac{1}{2}\left(|k+1|^{2H} - 2|k|^{2H} + |k-1|^{2H}\right)$

(3)

following a Hurst exponent $H$ can cause the moments of the autocorrelation to diverge for $0.5 < H_{fGn} \leq 1$ [38]. Curiously or confusingly, fGn is also known as "pink noise" [39], a usage that has defined how biologists and psychologists have begun to describe fGn-like structure in their measurements [40]. Though it could seem alien or cumbersome to a physics audience, we retain this usage here as a foothold for biologists and psychologists interested in joining the ergodicity discourse.

Critical to that ergodicity discourse, power-law decay of the autocorrelation is sufficient to make fGn nonergodic. And it is not simply that raw fGn series are nonergodic. While trying to model the variability of fGn, we might try to coarse-grain the nonergodicity with summary descriptors. However, the standard linear descriptors of variability are not guaranteed to damp out the ergodicity at all. For instance, they may easily inherit the nonergodicity of the raw fGn series. For instance, standard deviation (*SD*) of fGn is ergodic when applied to long samples of fGn—in this case, the stationarity of fGn might bolster the ergodicity of its *SD*, but descriptors that incorporate the mean such as coefficient of variation (*CV*) and root mean square (*RMS*) reliably break ergodicity as well as fGn [41].

    A meaningful way forward is to consider that the causes we should be modeling to explain biological and psychological processes are known to break ergodicity. The loss of mixing and breaking of ergodicity may have much to do with interdependencies among factors unfolding at multiple spatial and temporal scales [42–45]. It could be that the scale-invariant shape of the power-law autocorrelation is a more stable parameter for explaining the observed variety [46]. Indeed, the exponent HfGn shows none of the ergodicity breaking of the fGn series. Although fGn is a linear process, one possible explanation for power-law scaling is the nonlinear interactions across scales in cascade processes known to generate intermittent, nonergodic behavior [47]. We can estimate the strength of cascade dynamics through multifractal geometry, specifically through estimating the width of the multifractal spectrum (Δ*a*) and then through a *t*-test comparing that multifractal-spectrum width to spectrum widths for linear surrogates ($t_{MF}$). If cascades have any hope of explaining nonergodic behavior, then we might expect at a minimum that Δ*a* and $t_{MF}$ should at least avoid breaking ergodicity. Sure enough, all three descriptors: $H_{fGn}$, Δ*a*, and $t_{MF}$, were ergodic, even when applied to describing fGn series [41]. Hence, we might begin to see a way to restore broken ergodicity to our causal models. Empirically, these descriptors have actually been useful predictors of dependent measures of living and thinking systems [45]. But even for those eventual cases where these fractal and multifractal descriptors might fail, we can take some solace in the fact that these descriptors satisfy the ergodic assumptions and could potentially gain entry to causal status.

### 1.5. The present study: Restoring ergodicity broken by non-Gaussianity

The present study raises the stakes by breaking more conditions for ergodicity. As noted above, mixing is only one aspect of ergodicity, which entails independent selections from a Gaussian distribution with a stable mean and always identical variance ("white noise"). Our prior work had only reduced mixing by simulating fGn temporal correlations with $H_{fGn}$ > 0.5 and reduced mixing with Gaussian noise only broke ergodicity for the raw simulated pink-noise series and two descriptors of pink-noise variability: coefficient of variation (*CV*) and root-mean-square (*RMS*) [41]. Reduced mixing was insufficient to break ergodicity in standard deviations (*SD*) of pink-noise series, except at very short timescales. The long-run stationarity of fGn is sufficient to keep *SD* ergodic.

    The present work aims at the question of what elaboration to fGn (i.e., "pink noise") could undermine the ergodicity of *SD*. Breaking the independence of selections from a Gaussian distribution to produce fGn breaks some of the ergodicity [41], and now, we test whether adding non-Gaussianity to fGn could further break ergodicity. We know that series with "colored" non-Gaussian noise can break ergodicity [48]. In the literature on stochastic processes, a noise's "color" refers to the relative dominance of frequencies in the Fourier power spectrum, leaning on an analogy between oscillations in the Fourier spectrum of measurable frequencies with oscillations in the electromagnetic spectrum [39,49]. White noise reflects equal power across all frequencies, and pink noise reflects greater dominance of the lower-frequency, longer-wavelength (i.e., "redder") oscillations, with slow diminution towards higher-frequency, shorter-wavelength (i.e., "blue-

er"/"more-violet") oscillations. Hence, we expect that the introduction of non-Gaussianity to pink noise should replicate prior evidence of ergodicity breaking.

### 1.5.1. Prediction 1: Adding non-Gaussianity to long-range correlations will break ergodicity in *SD*, not simply in *CV* and *RMS*, as for fractional Gaussian noise

We expect further that stronger non-Gaussianity will accentuate the ergodicity breaking in the case of fGn. However, to be clear, non-Gaussianity in and of itself need not be a guarantee of ergodicity breaking. Indeed, various lognormal processes themselves can be ergodic in the parameters of the lognormal probability distribution function [50–52]. However, non-Gaussianity can be a liability for breaking ergodicity [53]. We specifically propose that non-Gaussianity might exacerbate the capacity of reduced mixing to break ergodicity. Individually, either non-Gaussianity or temporal correlations alone may be unremarkable extensions of the linear model entailing outliers or delays, respectively, with no apparent link to anything other than homogeneity and ergodicity [20,54,55]. However, the blend of the two is more certain to generate the turbulent stochastic processes [56] that more reliably break ergodicity [57,58]. Hence, we expect all the same ergodicity breaking in pink noise, as well as in *CV* and *RMS* of pink noise that we found before, but we expect that stronger non-Gaussianity (encoded as higher values of a parameter $\lambda$ defined in Section 2.1) will lead to progressively more breaking of ergodicity in the *SD* description of pink noise. We propose that uncorrelated non-Gaussian noise is the ergodic case, and long-range correlated non-Gaussianity breaks ergodicity; whereas reducing mixing by adding long-range correlation in fGn only leads ergodicity to fray slightly, non-Gaussianity conspires with reduced mixing to break ergodicity more dramatically.

### 1.5.2. Prediction 2: Fractal and multifractal measures $H_{fGn}$, $\Delta \alpha$, and $t_{MF}$ will provide ergodic description of non-Gaussian noise with long-range correlations

Now, the turbulent dynamics are more than just the sum of the two symptoms discussed: fGn-like long-range correlations and non-Gaussianity. On the contrary, turbulent dynamics exhibit these symptoms due to a deeper source, namely, cascade-like interdependencies across space and time scales [47,59]. These cascade dynamics can generate both power-law scaling in the autocorrelation consistent with fGn, and non-Gaussianity with nonzero $\lambda$. Therefore, as noted above, a first pass at describing this across-scale interdependencies can be the estimate of the fractal Hurst exponent $H_{fGn}$. Additionally, a more thorough description of these processes should be available in the width of the multifractal spectrum and the *t*-test comparing this width with that for linear surrogates [60]. We expect that $H_{fGn}$, $\Delta \alpha$, and $t_{MF}$ will all be as ergodic for non-Gaussian pink noise as for non-Gaussian white noise, thereby restoring ergodicity where the blend of non-Gaussianity with long-range correlations had broken it.

## 2. Methods

### 2.1. Simulating non-Gaussian white and pink noises

If a stochastic process $\{x_i\}$ can be described as

$$x_i = \xi_i e^{\omega_i} \tag{4}$$

where $\xi$ is a Gaussian variable with zero mean and $\omega$ is also a Gaussian independent of $\xi$, the PDF of $\{x_i\}$, it is possible to assume that $\{\xi_i\}$ and $\{\omega_i\}$ are both uncorrelated random variables, although the long range correlation of the $\omega_i$ plays an important role in generating intermittency [61,62]. A stochastic process $\{x_i\}(i=1,2,3,...)$ exhibiting the non-Gaussian PDF of $f_\lambda(x)$ [Eq. (4)] is described by

$$x_i = \xi_i e^{\lambda \omega_i - \lambda^2} \tag{5}$$

where $\{\xi_i\}$ and $\{\omega_i\}$ are both sequences of Gaussian noise with zero mean and unit variance, and independent of each other; and $\lambda$ is the non-Gaussian parameter. The $x_i$ is a Gaussian variable when $\lambda=0$. Similarly, Eq. (5) converges to a Gaussian in the limit $\lambda \to 0$.

We simulated 100 50,000-samples series of non-Gaussian white and pink noises [i.e., by plugging two independent white noise and two independent pink noise series in place of $\{\xi_i\}$ and $\{\omega_i\}$ in Eq. (5)] for the following values of $\lambda$: $\lambda=0.1,0.2,0.3,0.4,0.5,0.6,0.7,0.8,0.9,1$ (figure 2), using MATLAB (Matlab Inc, Natick, MA). All series were then unsigned and all analysis was conducted on these unsigned values. Using unsigned values is a common practice in fractal and multifractal analysis. A shuffled version of each original non-Gaussian white noise and pink noise series was generated for comparison, specifically because ergodicity is about how sequence exemplifies a typical mean trajectory of a sample of realizations. By breaking the sequence, shuffling of all non-Gaissian noises produces non-Gaussian white-noise fluctuations around the mean.

## 2.2. Estimating linear descriptors

We computed the following linear indices over nonoverlapping epochs of 250, 500, 100, and 2000 samples—amounting to 200, 100, 50, and 25 epochs, respectively—for the original version (i.e., unshuffled) and a shuffled version (i.e., a version with the temporal information destroyed) of each pink and white noise series:

Standard deviation (*SD*), that is,

$$\sqrt{\frac{1}{N} \sum_{i=1}^{N} (x_i - \mu)^2}, \tag{6}$$

where $N$ is sample size (i.e., 250, 500, 100, or 2000), $x_i$ is $i$th individual value, and $\mu$ is sample mean.

Coefficient of variation (*CV*), that is,

$$\frac{SD}{\mu}, \tag{7}$$

where $\mu$ is sample mean.

Root mean square (*RMS*), that is,

$$\sqrt{\frac{1}{N} \sum_{i=1}^{N} |x_i|^2}, \tag{8}$$

where where $N$ is sample size, $x_i$ is $i$th individual value, and $\mu$ is sample mean.

## 2.3. Fractal and multifractal descriptors

### 2.3.1. Accessing fractality using detrended fluctuation analysis

Detrended fluctuation analysis (DFA) computes Hurst exponent $H_{fGn}$ quantifying temporal correlations in white noise and pink noise series [63,64] using the first-order integration of $N$-length time series $x(t)$:

$$y(t) = \sum_{i=1}^{N} x(t) - \overline{x(t)}, \qquad (9)$$

where $\overline{x(t)}$ is the time-series grand mean. It computes root mean square (RMS; i.e., averaging the residuals) for each linear trend $y_n(t)$ fit to nonoverlapping $n$-length bins ( ) to build fluctuation function $f(\ )$:

$$f(N) = \sqrt{(1/N) \sum_{i=1}^{N} \left(x(t) - \overline{x(t)}\right)^2}, \qquad (10)$$

for $n < N/4$. On standard scales, $f(N)$ is a power law:

$$f(N) \sim {}^{\square} n^H, \qquad (11)$$

where $H_{fGn}$ is the scaling exponent estimable using logarithmic transformation:

$$\log f(N) = H_{fGN} \log(n). \qquad (12)$$

Higher $H_{fGn}$ corresponds to stronger temporal correlations. We computed $H_{fGn}$ over 250-, 500-, 100-, and 2000-sample epochs for the original and shuffled unsigned white noise and pink noise series.

### 2.3.2. Multifractal analysis

### 2.3.2.1. Assessing multifractal nonlinearity using the direct-estimation of singularity spectrum

Chhabra and Jensen's [65] direct method estimated multifractal spectrum widths $\Delta \alpha$ for each 250-, 500-, 100-, and 2000-samples epoch for the original and shuffled unsigned pink noise and white noise series. This method samples series $u(t)$ at progressively larger scales using the proportion of signal $P_i(L)$ falling within the $i^{th}$ bin of scale $L$ is

$$P_i(L) = \frac{\sum_{k=(i-1)L+1}^{iL} u(k)}{\sum u(t)}. \qquad (13)$$

As $L$ increases, $P_i(L)$ represents progressively larger proportion of $u(t)$,

$$P(L) \propto L^{\alpha} \qquad (14)$$

suggesting growth of proportion according to one "singularity" strength $\alpha$ [66]. $P(L)$ exhibits multifractal dynamics when it grows heterogeneously across time scales $L$ according to multiple singularity strengths, such that

$$P_i(L) \propto L^{\alpha_i} \qquad (15)$$

whereby each $i^{th}$ bin may show a distinct relationship of $P(L)$ with $L$. The width of this singularity spectrum, $\Delta \alpha (\alpha_{max} - \alpha_{min})$, indicates the heterogeneity of these relationships [67,68].

Chhabra and Jensen's method [65] estimates $P(L)$ for $N_L$ nonoverlapping bins of $L$-sizes and transforms them into a "mass" $\mu(q)$ using a $q$ parameter emphasizing higher or lower $P(L)$ for $q>1$ and $q<1$, respectively, as follows

$$\mu_i(q,L) = \frac{[P_i(L)]^q}{\sum_{i=1}^{N_L}[P_i(L)]^q}. \tag{16}$$

$\alpha(q)$ is the singularity for mass $\mu$-weighted $P(L)$ estimated by

$$\alpha(q) = -\lim_{L \to \infty} \frac{1}{\ln L} \sum_{i=1}^{N} \mu_i(q,L) \ln P_i(L)$$

$$¿\lim_{L \to 0} \frac{1}{\ln L} \sum_{i=1}^{N} \mu_i(q,L) \ln P_i(L). \tag{17}$$

Each estimated value of $\alpha(q)$ belongs to the multifractal spectrum only when the Shannon entropy of $\mu(q,l)$ scales with $L$ according to the Hausdorff dimension $f(q)$, where

$$f(q) = -\lim_{L \to \infty} \frac{1}{\ln L} \sum_{i=1}^{N} \mu_i(q,L) \ln \mu_i(q,L)$$

$$¿\lim_{L \to 0} \frac{1}{\ln L} \sum_{i=1}^{N} \mu_i(q,L) \ln \mu_i(q,L). \tag{18}$$

For values of $q$ yielding a strong relationship between Eqs. (17 & 18)—in this study, $r > 0.995$, the parametric curve $(\alpha(q), f(q))$ or $(\alpha, f(\alpha))$ constitutes the multifractal spectrum. We obtained the multifractal-spectrum width $\Delta\alpha$ over nonoverlapping epochs of 250, 500, 100, and 2000 samples for the original and shuffled versions of each white noise and pink noise series.

### 2.3.2.2. Surrogate testing using Iterated Amplitude Adjusted Fourier Transformation (IAAFT) generated $t_{MF}$

To identify whether nonzero $\Delta\alpha$ reflected multifractality due to nonlinear interactions across timescales, $\Delta\alpha$ for the original and a shuffled version of each pink and white noise series was compared to $\Delta\alpha$ for 32 IAAFT surrogates [69,70]. IAAFT randomizes original values time-symmetrically around the autoregressive structure, generating surrogates that randomized phase ordering while preserving linear temporal correlations. The one-sample $t$-statistic (henceforth, $t_{MF}$) takes the subtractive difference between $\Delta\alpha$ for the original series and that for the 32 surrogates, dividing by the standard error of the spectrum width for the surrogates. We obtained $t_{MF}$ computed over nonoverlapping epochs of 250, 500, 100, and 2000 samples for the original and shuffled versions of each white noise and pink noise series.

### 2.5. Estimating ergodicity breaking in raw and SD, CV, RMS, $H_{fGn}$, $\Delta a$, and $t_{MF}$ series

Ergodicity can be quantified using a dimensionless statistic of ergodicity breaking $E_B$ — also known as the Thirumalai-Mountain metric [71], computed by subtracting squared the total-sample variance from the average squared subsample variance and dividing the resultant by the total-sample squared variance.

$$E_B(x(t)) = \frac{\left(\left\langle\left[\delta^2(x(t))\right]^2\right\rangle - \left\langle\overline{\delta^2}(x(t))\right\rangle\right)}{\left\langle\overline{\delta^2}(x(t))\right\rangle^2} \tag{19}$$

Rapid convergence of $E_B$ for progressively larger samples, that is, $E_B \to 0$ as $t \to \infty$ implies ergodicity. Slower convergence indicates weaker ergodicity, and no convergence indicates nonergodicity [10]. $E_B$ thus serves a simple way to test whether a given statistical descriptor fulfills ergodic assumptions or breaks ergodicity and compare nonergodicity of one statistical descriptor to that of the other. We computed the ergodicity breaking factor $E_B$ for each original and shuffled raw unsigned white noise and pink noise series, for lag $\Delta$ = 2 samples, and for each SD, CV, RMS, $H_{fGn}$, $\Delta a$, and $t_{MF}$ series computed over nonoverlapping epochs of 250, 500, 100, and 2000 samples each for the original and shuffled unsigned white noise and pink noise, for $\Delta$ = 2 epochs.

## 3. Results

### 3.1. Temporal correlations but not non-Gaussianity lead to weak mixing and ergodicity breaking

The unsigned pink-noise series shows considerably higher variability than the unsigned white-noise series, and the series' non-Gaussianity seemed to accentuate this difference. Even a cursory examination of figures 3a and 3b reveals these trends, reflecting possible ergodicity-related differences across the two noises and levels of non-Gaussianity: (i) white noise is more likely than unsigned pink-noise series to return and, over a bigger ensemble, converge towards the mean, and is thus more ergodic than pink noise; and (ii) series with low non-Gaussianity is more likely than series with high non-Gaussianity to return and, over a bigger ensemble, converge towards the mean, and is thus more ergodic than series with high non-Gaussianity. The $E_B$ vs. epochs curves, which almost entirely coincide for the original and shuffled unsigned white-noise series but do not coincide for the original and shuffled unsigned pink-noise series, demonstrate this nonergodic behavior of pink noise (figures 3c and 3d). Although the $E_B$ vs. epochs curve for the shuffled unsigned pink-noise series was no different than the $E_B$ vs. epochs curves for the original and shuffled unsigned white-noise series, the original unsigned pink-noise series exhibited ergodicity breaking with shallower $E_B$ curves. The $E_B$ vs. epochs curve detect non-Gaussianity, as an increase in non-Gaussianity in the white-noise series leads to an upward shift of the $E_B$ vs. $t$ curves, but in the absence of temporal correlations in white-noise-series, non-Gaussianity was not enough to break ergodicity. Notably, the increase in non-Gaussianity of pink-noise series was associated with both an upward shift and change in shape of the $E_B$ vs. $t$ curves, indicating that temporal correlations and non-Gaussianity break ergodicity in nuanced ways. Hence, pink noise is only faintly mixing and breaks ergodicity, and injecting non-Gaussianity in pink noise influences its ergodicity breaking behavior.

### 3.2. Confirming Prediction 1: Linear statistics such as *SD*, *CV*, and *RMS* break ergodicity

Whereas long-range correlations does not reliably break ergodicity of *SD* in Gaussian noise [41], non-Gaussianity work together with long-range correlations to break SD. There is no difference between SD series for the original and shuffled unsigned white noise, although *SD* values are marginally higher for series with higher non-Gaussianity (figure 4a). In contrast, *SD* series for the original unsigned pink noise is consistently higher (figure 4b), further confirming that variability grows faster in pink noise than white noise. Although *SD* series for the original and shuffled unsigned white noise does not differ in their ergodicity breaking behavior (figure 4c), SD series for the original unsigned pink noise show weak but marginally stronger ergodicity breaking behavior than SD series for the shuffled unsigned pink noise (figure 4d). Notably, the increase in non-Gaussianity is associated with both an upward shift and change in shape of the $E_B$ vs. epochs curves for SD series of both pink and white noises. As $\lambda$ increases from 0 to 1, the $E_B$ vs. epochs curves for SD series of the original unsigned pink noise become progressively shallower, diverging more so from the $E_B$ vs. epochs curves for SD series of the shuffled unsigned pink noise (Figure 4e).

Although no difference exists between CV series for the original and shuffled unsigned white noise (figure 5a), CV series for the original unsigned white noise is consistently higher than for the original unsigned pink noise (figure 5b). Likewise, CV series for the original and shuffled unsigned white noise does not differ in their ergodicity breaking behavior (figures 5c), but CV series show weak ergodicity breaking for the original unsigned pink noise compared to the shuffled unsigned pink noise, as indicated by shallower $E_B$ vs. epochs curves (figures 5d). Not surprisingly, the ergodicity breaking behavior of CV series for the original unsigned pink noise does not depend on the series' non-Gaussianity, possibility due to the proportional growths of variability and mean in pink noise.

RMS series behave pretty similar to SD series except that RMS series for the original unsigned pink noise (figure 6b) is consistently higher than for the original unsigned white noise (figure 6a), confirming that variability grows faster in pink noise than white noise. Additionally, RMS series for the original and shuffled unsigned white noise does not differ in their ergodicity breaking behavior (figures 6c), but RMS series show weak ergodicity breaking for the original unsigned pink noise compared to the shuffled unsigned pink noise, as indicated by shallower $E_B$ vs. epochs curves (figures 6d). Notably, the increase in non-Gaussianity was associated with both an upward shift and change in shape of the $E_B$ vs. epochs curves for RMS series of both pink and white noises. Overall, independent of the level of non-Gaussianity, SD, CV, and RMS series all show weak ergodicity breaking for the original unsigned pink noise. In other words, SD, CV, and RMS fail to encode the nonergodicity arising from temporal correlations inherent in pink noise.

### 3.3. Confirming Prediction 2: Fractal and multifractal statistics such as $H_{fGn}$, $\Delta\alpha$, and $t_{MF}$ fulfill the ergodic assumption independent of non-Gaussianity

We next investigate the ergodicity breaking behavior of $H_{fGn}$, $\Delta\alpha$, and $t_{MF}$ series for the white and pink noise with different levels of non-Gaussianity, using the Thirumalai-Mountain method [71], with each of these statistical descriptors computed across 50 epochs of 1000-samples each. $H_{fGn}$ series neither differ between the original and shuffled unsigned white noise nor differ across series with different levels of non-Gaussianity (figure 7a). In contrast, although $H_{fGn}$ series for the original and shuffled unsigned pink noise does not differ across series with different levels of non-Gaussianity, $H_{fGn}$ series for the original unsigned pink noise show consistently higher and more variable values than for the original unsigned white noise (figure 7b). Notably, $E_B \rightarrow 0$ as $t \rightarrow \infty$ for the original and shuffled unsigned white noise—conifrming that $H_{fGn}$ series for white noise is ergodic, the $E_B$ vs. epochs curves for the original and shuffled unsigned white noise are marginally higher for series' with higher non-Gaussianity (figure 7c). In contrast, while $E_B \rightarrow 0$ as $t \rightarrow \infty$ for the original and shuffled unsigned white noise and pink noise—conifrming that $H_{fGn}$ series for pink noise is also ergodic, the $E_B$ vs. epochs curves for the original and unsigned unsigned pink noise does not at all depend on the series' non-Gaussianity (figure 7d).

Much like for $H_{fGn}$ series, $\Delta a$ series take comparable values for the original and shuffled unsigned white noise but greater and consistently more variable values for the original unsigned pink noise than for the shuffled unsigned pink noise or white noise (figures 8a and 8b). Despite this higher variability in $\Delta a$ values observed for the original unsigned pink noise, the $E_B$ vs. epochs curves scarcely differ between $\Delta a$ series for the original and shuffled unsigned white noise and for the original and shuffled unsigned pink noise (figures 8c and 8d). Indeed, $E_B \to 0$ as $t \to \infty$ for the original and shuffled unsigned pink noise white noise—conifrming that $\Delta a$ series for white noise and pink noise is also ergodic, and the $E_B$ vs. epochs curves for the original and unsigned unsigned pink noise does not at all depend on the series' non-Gaussianity (figures 8c and 8d). $t_{MF}$ series behaved similar to $\Delta a$ series both in terms of the values that it take (figures 9a and 9b) and how the associated $E_B$ vs. epochs curves span out (figures 9c and 9d). So, both $\Delta a$ and $t_{MF}$ series respect the ergodic assumption.

In summary, fractal and multifractal statistics such as $H_{fGn}$, $\Delta a$, and $t_{MF}$ respect the ergodic assumption independent of non-Gaussianity (at least in the range $\lambda = 0.1$–$1$). These fractal and multifractal statistics are themselves ergodic while explicltly encoding the nonergodicity from cascade-like dynamics in long-range correlated non-Gaussian noise. These fractal and multifractal statistics themselves encode the degree and type of nonergodicity in measurement series.

### 3.4. Effects of epoch length on ergodicity breaking

The $E_B$ vs. epochs curves for the original and shuffled unsigned white noise observed in figures 4–7 are highly sensitive to the setting of epoch size, for SD series (figures 10a and 10b), CV series (figures 10c and 10d), and RMS series (figure 10e and 10f), and this sensitivity is moderated by non-Gaussianity. The $E_B$ vs. epochs curves for white noise does not depend on epoch size for low non-Gaussianity but increasingly depended on epoch size for high non-Gaussianity. The $E_B$ vs. epochs curves for pink noise depended on epoch size for both low and high non-Gaussianity with this dependence being more prominent for high non-Gaussianity. Overall, these trends suggest that the SD, CV, and RMS series break ergodicity for both white and pink noise, irrespective of the series' nonergodicity. Notably, this ergodicity breaking behavior depends on the chosen epoch size.

The $E_B$ vs. epochs curves for the original and shuffled unsigned white noise and pink noise observed in figures 7–9 are largely insensitive to the setting of epoch size, for $H_{fGn}$ series (figures 11a and 11b), $\Delta a$ series (figures 11c and 11d), and $t_{MF}$ series (figures 11e and 11f), independent of the series' non-Gaussianity. Hence, while the linear descriptors based on quantifying the second moment show ergodicity breaking, nonlinear descriptors quantifying fractal and multifractal fluctuations show ergodic behavior that neither depends on the series' non-Gaussianity nor the epoch size used for the analysis. Minor exceptions included the $E_B$ vs. epochs curves for $\Delta a$, which show greater dependence on epoch size for very high non-Gaussianity in white noise, and the $E_B$ vs. epochs curves, which are higher for very high non-Gaussianity ($\lambda = 1$; dark red) than very low non-Gaussianity ($\lambda = 0.1$; dark blue) in both white noise and pink noise. Otherwise, the trends suggest that $H_{fGn}$, $\Delta a$, and $t_{MF}$ series for both white noise and pink noise are ergodic independent of the series' non-Gaussianity and the chosen epoch size.

### 4. Discussion

The present work continues earlier simulation work and allows us to develop quantitative descriptors of nonergodic processes that will serve biological and psychological approaches to causal modeling. This work sheds more light on the challenges facing any field in which the measurements exhibit fGn-like long-range correlations. The "pink noise" measurement series that we have found pervading across biological and psychological processes appears to be a crucial threat to ergodicity. Greater non-Gaussianity increases this threat of pink noise to our causal modeling. It may be a natural and intuitive impulse to attempt to quantify the variability in our measures using familiar linear summary descriptors like *SD*, *CV*, and *RMS*, but these descriptors fail to restore any ergodicity to our causal modeling. Among those three measures, *SD* was found to be relatively robust to ergodicity breaking [41], but the present work highlights the limitatitons of that robustness. The present work suggests that *SD* is robust to non-Gaussianity alone, and our previous work suggested that *SD* is also robust to long-range correlations in the long run. However, the present work indicates that non-Gaussianity and temporal correlations break ergodicity in *SD*. This point is alarming given the prevalence of these features in measurements of biological and psychological processes [15,45].

The present work makes a compelling case, even in this perfect storm of broken ergodicity, we can find ergodicity in statistical descriptors that aim to estimate the cascade dynamics known to generate this storm. Hence, a long-range correlated non-Gaussian process in our measurements may break ergodicity, and linear summary descriptors of such a nonergodic process will also break ergodicity. These linear descriptors may reflect an attempt to shoehorn unfamiliar stochastic processes into familiar statistical frameworks. However, including statistical descriptors that make explicit reference to plausible generating mechanisms gives us the chance to find ergodic descriptors that could operationalize possible causes to give our linear causal models a chance to stand the test. This theoretical work bolsters ongoing empirical research that has consistently found that fractal and multifractal descriptors of measurement series offer a set of compelling, statistically significant predictors of behavioral and cognitive performance [32–37,45,72–74].

**Data accessibility.** This article has no additional data.

**Author contributions.** D.G.K-S. and M.M. conceived and designed research; M.M. performed simulations; D.G.K-S. and M.M. analyzed simulation results; D.G.K-S. and M.M. interpreted simulation results; M.M. prepared figures; D.G.K-S. and M.M. drafted manuscript; D.G.K-S. and M.M. edited and revised manuscript; D.G.K-S. and M.M. approved final version of manuscript.

**Competing interests.** We declare we have no competing interests.

**Funding.** We received no funding for this work.

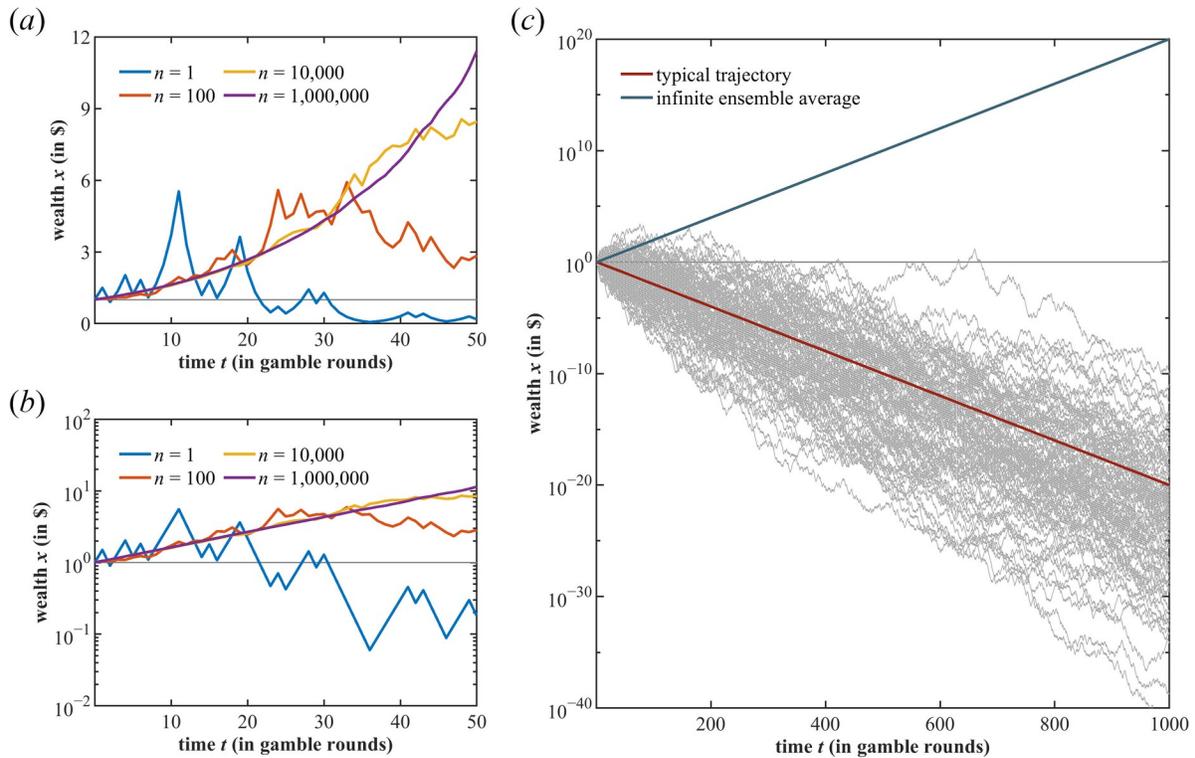

**Figure 1.** Example nonergodic process: repeated gambling. A repeated gambling process is modeled as follows: the gambler tosses a coin; for heads, they win 50% of their current wealth of $1; for tails, they lose 40% of their current wealth. (*a*) A typical gambler will always go bust after 50 gambling rounds. The average wealth of 100 concurrent gamblers shows no trend (red line), the average wealth of 10,000 concurrent gamblers appears to harvest a nice ~10x profit (yellow line), and the average wealth of 1,000,000 concurrent gamblers show a similar win (purple line). (*b*) A logarithmic *y*-axis shows the simulation in (*a*) more clearly. (*c*) The average wealth of an infinite ensemble of gamblers (blue line) portrays a different picture than the wealth of a typical gambler over time (red line). The illustration shows wealth trajectories of 100 gamblers across 1000 gambling rounds. Adapted from [6].

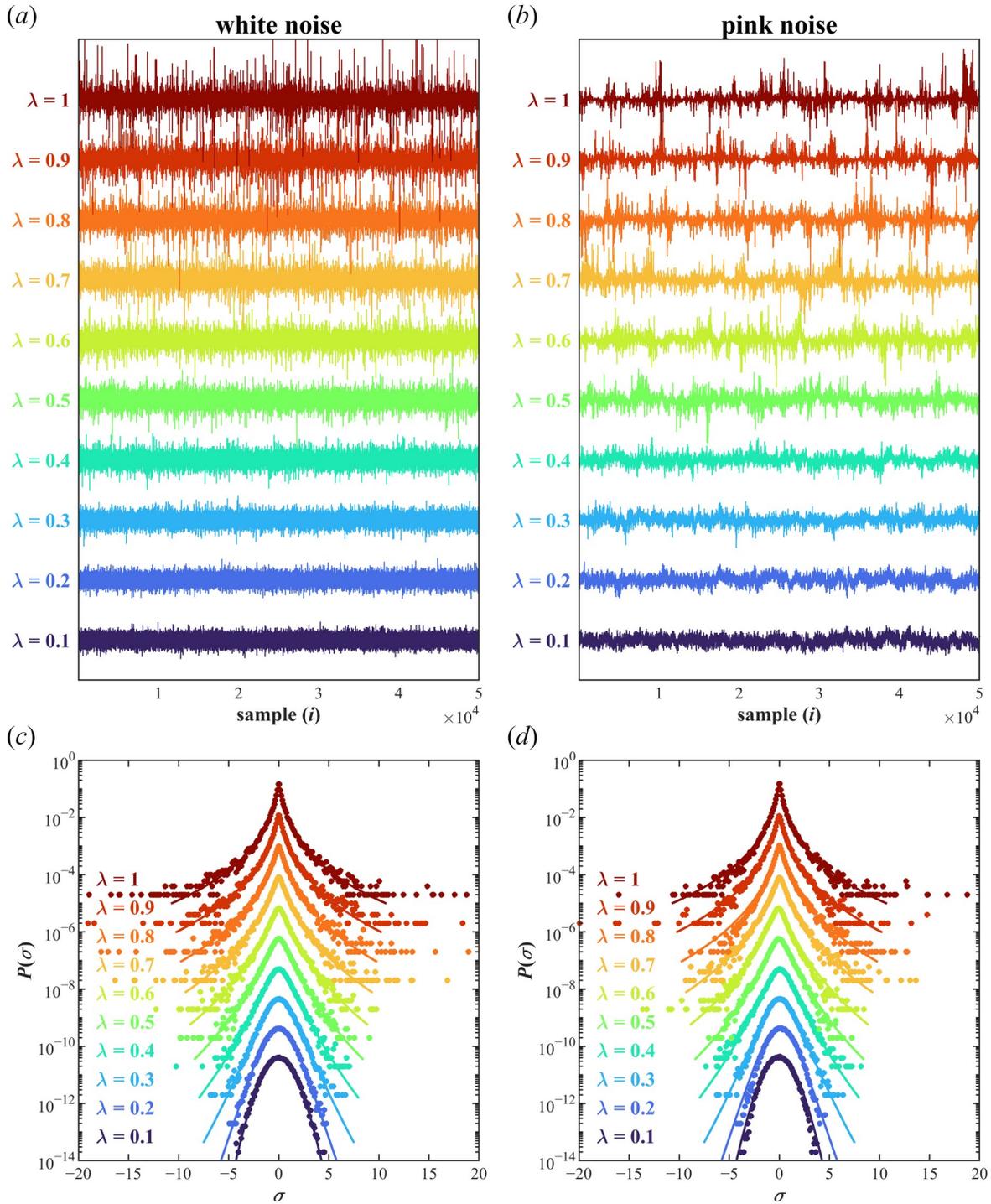

**Figure 2.** Simulated white and pink noises with different levels of non-Gaussianity. (*a*) Time series of white noise with different levels of non-Gaussanity, quantified by the index $\lambda$, as described by Eq. (6). (*b*) Time series of pink noise with different levels of non-Gaussanity, quantified by the index $\lambda$, as described by Eq. (7). (*c*) Probability density functions (PDFs) of white noise in (*a*). (*d*) PDFs of pink noise in (*b*). Solid lines indicate the numerical integration of Eq. (3). Symbols indicate the estimated PDF's from the simulated white and pink noises shown in (*a*) and (*b*). The PDFs have been shifted vertically for convenience of presentation, and hence, the vertical axis is given in arbitrary units.

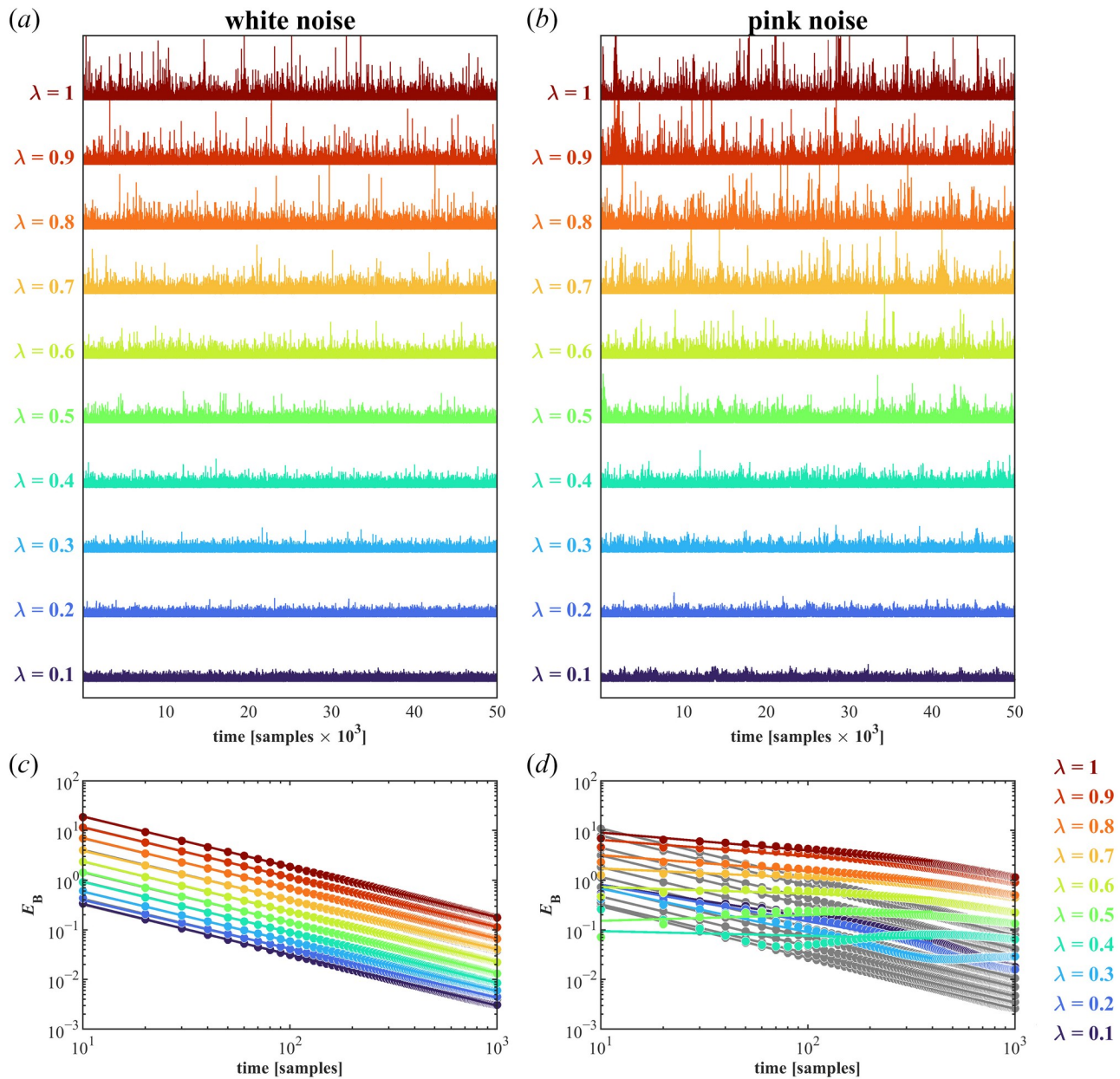

**Figure 3.** Ergodicity breaking in unsigned white noise and pink noise series with different levels of non-Gaussianity. (*a*) Simulated unsigned white-noise series with different levels of non-Gaussianity. (*b*) Simulated unsigned pink-noise series with different levels of non-Gaussianity. A noteworthy difference between (*a*) and (*b*) is that the unsigned pink noise shows marginally more variability than the unsigned white noise at all levels of non-Gaussianity. So, even just visual inspection provides an early sense of a potential difference in ergodicity: unsigned white noise appears relatively more likely than unsigned pink noise to return and, over a larger ensemble, converges towards the mean. (*c*) and (*d*) show how the ergodicity-breaking parameter ($E_B$; for lag $\Delta = 2$ samples) differs between the original unsigned white- and pink-noise series for different levels of non-Gaussianity and their shuffled versions. This comparison is critical because ergodicity is fundamentally about the sequence—how the sequence of a series exemplifies a typical mean trajectory of a sample of comparable systems. Shuffling breaks the sequence and guarantees that the series' trajectory fluctuates around the mean. (c) $E_B$ vs. time *t* for the original unsigned white-noise series (*mean±s.d. $H_{fGn}$* = 0.4931±0.0034) and its shuffled counterparts ($H_{fGn}$ = 0.4957±0.0037). Because unsigned white noise has no temporal correlations, shuffling an unsigned white-noise series produces another series of unsigned white noise. Hence, the two $E_B$ vs. $t$ curves coincide almost completely, with the grey shuffled-series' curve scarcely visible with a greater difference in the right half. An increase in non-Gaussianity in the white-noise series leads to an upward shift of the $E_B$ vs. $t$ curves. (*d*) $E_B$ vs. $t$ for the original unsigned pink-noise series ($H_{fGn}$ = 0.6849±0.0041) and its shuffled counterparts ($H_{fGn}$ = 0.4955±0.0021). Shuffling an unsigned pink-noise series produces an unsigned white-noise series due to loss of temporal correlations, illustrated by the fact that the grey curve for the shuffled unsigned pink-noise series resembles both curves in (*c*). $H_{fGn} \to 1$ yields nonergodic behavior as indicated by the higher and, for all but the largest sample size, shallower $E_B$ curve [10]. Notably, the increase in non-Gaussianity of pink-noise series was associated with both an upward shift and change in shape of the $E_B$ vs. $t$ curves, indicating that temporal correlations and non-Gaussianity break ergodicity in nuanced ways.

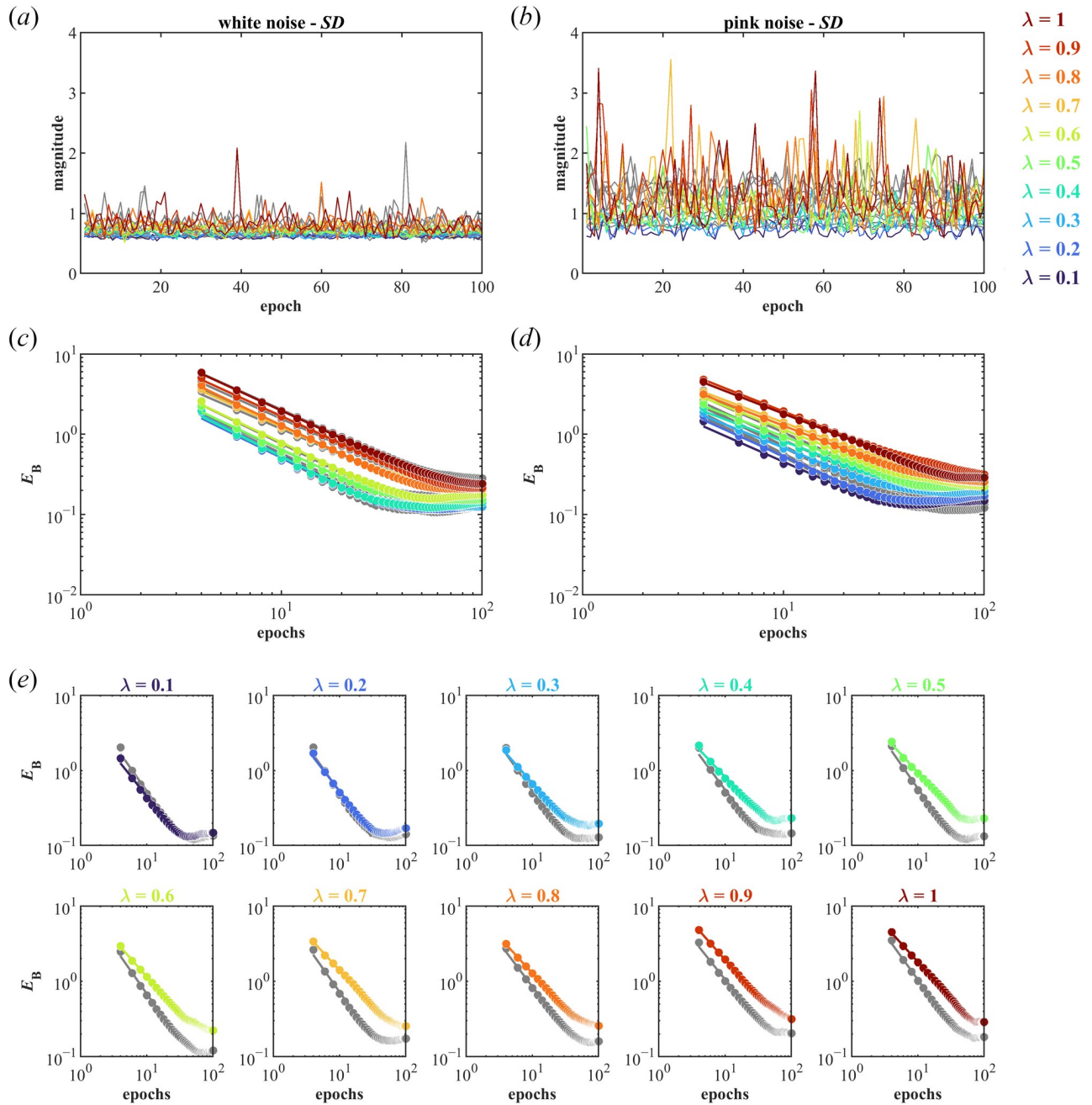

**Figure 4.** *SD* series for the original unsigned pink noise breaks ergodicity depending on the level of non-Gaussianity. Ergodicity breaking in *SD* series (*SD* calculated across 500-samples epochs) for unsigned white noise (*a*), and unsigned pink noise (*b*), with different levels of non-Gaussianity. *SD* series for the original unsigned pink noise is higher than the *SD* series for the shuffled unsigned pink noise, and non-Gaussianity accentuated this trend, but no difference exists between *SD* series for the original and shuffled unsigned white noise. The overall greater *SD* in (*b*) than (*a*) reflects that variability can grow faster in unsigned pink noise than unsigned white noise. $E_B$ vs. epochs (for lag $\Delta = 2$ epochs) for *SD* series for the original and shuffled unsigned white noise (*c*), and the original and shuffled unsigned pink noise (*d*), with different levels of non-Gaussianity. $E_B$ for *SD* series decreases steadily with epochs and then levels off at larger values at epochs. A relatively slower decay in $E_B$ with epochs is observed for *SD* series for the original than shuffled unsigned pink noise, but no such distinction is observed between *SD* series for the original and shuffled unsigned white noise. The $E_B$ vs. epochs curves are increasingly higher for higher non-Gaussianity (dark blue through dark red curves for $\lambda = 0.1–1.0$) in both white noise and pink noise. (e) The $E_B$ vs. epochs curves for the original unsigned pink noise become progressively shallower as $\lambda$ increases.

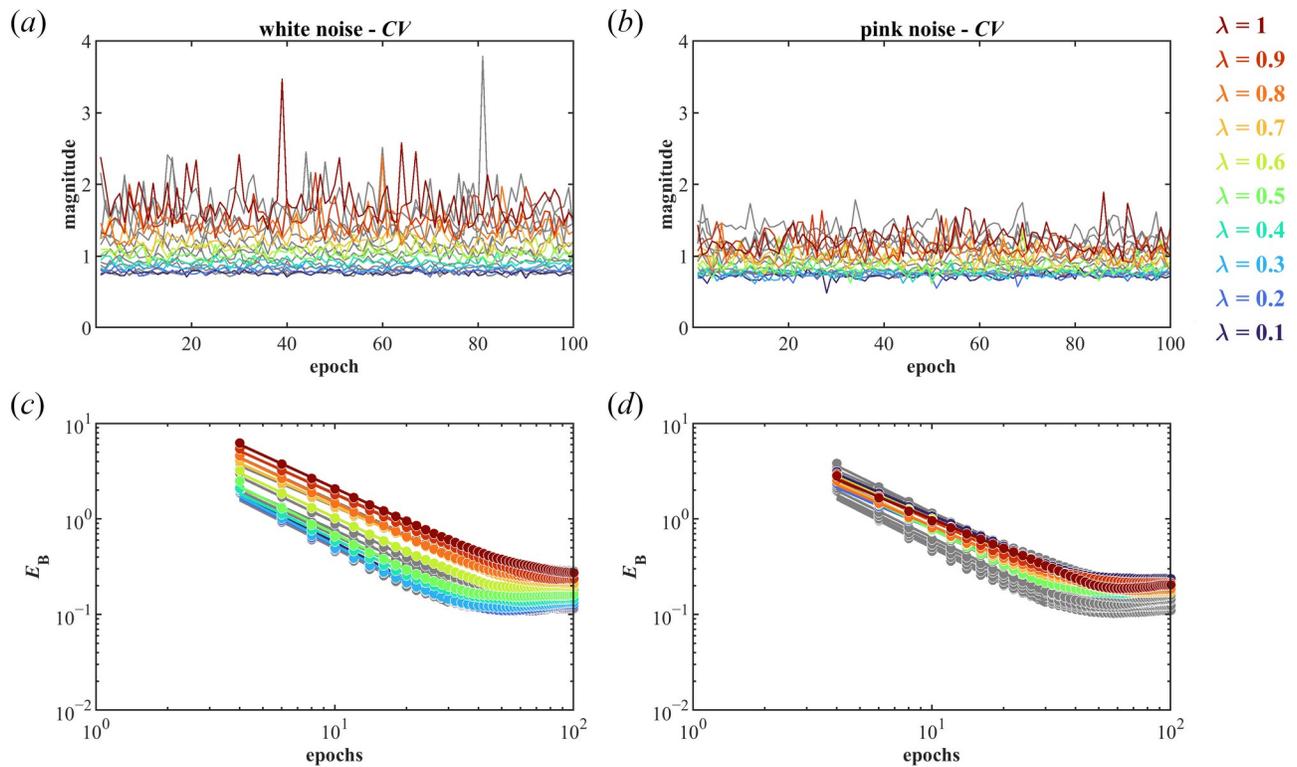

**Figure 5.** *CV* series for the original unsigned pink noise breaks ergodicity depending on the level of non-Gaussianity. Ergodicity breaking in *CV* series (*CV* calculated across 500-samples epochs) for unsigned white noise (*a*), and unsigned pink noise (*b*), with different levels of non-Gaussianity. *CV* series for the original unsigned pink noise is higher than the *CV* series for the shuffled unsigned pink noise, and non-Gaussianity accentuated this trend, but no difference exists between *CV* series for the original and shuffled unsigned white noise. The overall greater *CV* in (*a*) than (*b*) reflects the fact that although variability grows faster in unsigned pink noise than unsigned white noise, this growth of variability in unsigned pink noise is proportional to the growth of the mean—a greater mean for unsigned pink noise yielded smaller *CV* values. $E_B$ vs. epochs (for lag $\Delta = 2$ epochs) for *CV* series for the original and shuffled unsigned white noise (*c*), and the original and shuffled unsigned pink noise (*d*), with different levels of non-Gaussianity. $E_B$ for *CV* series decreases steadily with epochs and then levels off at larger values at epochs. A relatively slower decay in $E_B$ with epochs is observed for *CV* series for the original than shuffled unsigned pink noise, but no such distinction is observed between *CV* series for the original and shuffled unsigned white noise. The $E_B$ vs. epochs curves are increasingly higher for higher non-Gaussianity (dark blue through dark red curves for $\lambda = 0.1$–$1.0$) in both white noise and pink noise.

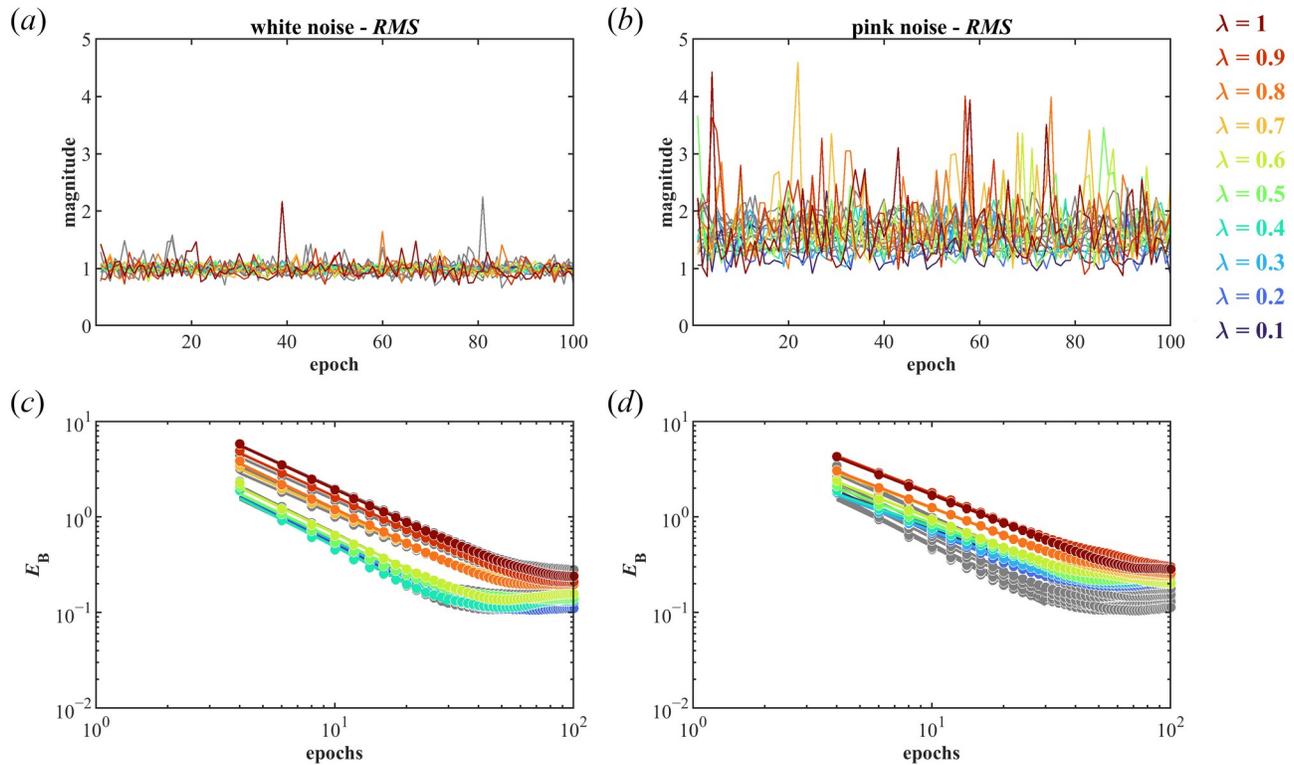

**Figure 6.** *RMS* series for the original unsigned pink noise breaks ergodicity depending on the level of non-Gaussianity. Ergodicity breaking in *RMS* series (*RMS* calculated across 500-samples epochs) for unsigned white noise (*a*), and unsigned pink noise (*b*), with different levels of non-Gaussianity. *RMS* series for the original unsigned pink noise is higher than the *RMS* series for the shuffled unsigned pink noise, and non-Gaussianity accentuated this trend, but no difference exists between *RMS* series for the original and shuffled unsigned white noise. The overall greater *RMS* in (*b*) than (*a*) reflects that variability can grow faster in unsigned pink noise than unsigned white noise. $E_B$ vs. epochs (for lag $\Delta = 2$ epochs) for *RMS* series for the original and shuffled unsigned white noise (*c*), and the original and shuffled unsigned pink noise (*d*), with different levels of non-Gaussianity. $E_B$ for *RMS* series decreases steadily with epochs and then levels off at larger values at epochs. A relatively slower decay in $E_B$ with epochs is observed for *RMS* series for the original than shuffled unsigned pink noise, but no such distinction is observed between *RMS* series for the original and shuffled unsigned white noise. The $E_B$ vs. epochs curves are increasingly higher for higher non-Gaussianity (dark blue through dark red curves for $\lambda = 0.1$ through 1.0) in both white noise and pink noise.

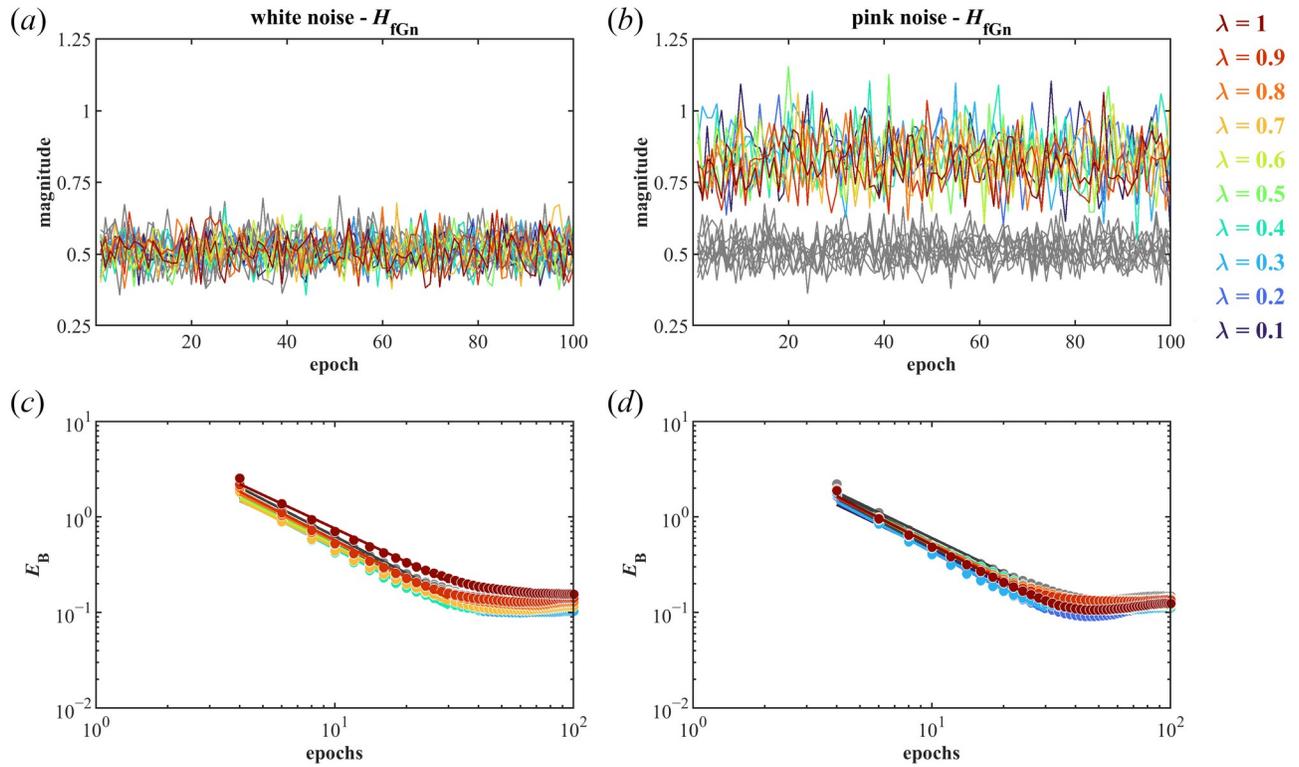

**Figure 7.** $H_{fGn}$ series for the original unsigned pink noise is ergodic irrespective of the level of non-Gaussianity. Ergodicity breaking in $H_{fGn}$ series ($H_{fGn}$ calculated across 500-samples epochs) for unsigned white noise (*a*), and unsigned pink noise (*b*), with different levels of non-Gaussianity. $H_{fGn}$ series for the original unsigned pink noise is higher than the $H_{fGn}$ series for the shuffled unsigned pink noise due to the presence of temporal correlations in the original unsigned pink noise, but no difference exists between $H_{fGn}$ series for the original and shuffled unsigned white noise. Notably, $H_{fGn}$ does not depend on the series' non-Gaussianity. $E_B$ vs. epochs (for lag $\Delta = 2$ epochs) for $H_{fGn}$ series for the original and shuffled unsigned white noise (*c*), and the original and shuffled unsigned pink noise (*d*), with different levels of non-Gaussianity. $E_B$ for $H_{fGn}$ series decreases steadily with epochs and then levels off at larger values at epochs. What emerges is that $H_{fGn}$ series shows none of the ergodicity breaking that the raw series for the original unsigned pink noise does in figure 3. The decay in $E_B$ vs. epochs only differs slightly between $H_{fGn}$ series for the original and shuffled unsigned white noise. It shows almost no difference between $H_{fGn}$ series for the original and shuffled unsigned pink noise. Notably, $E_B$ vs. epochs curves for $H_{fGn}$ series for the original unsigned pink noise does not depend on the series' non-Gaussianity.

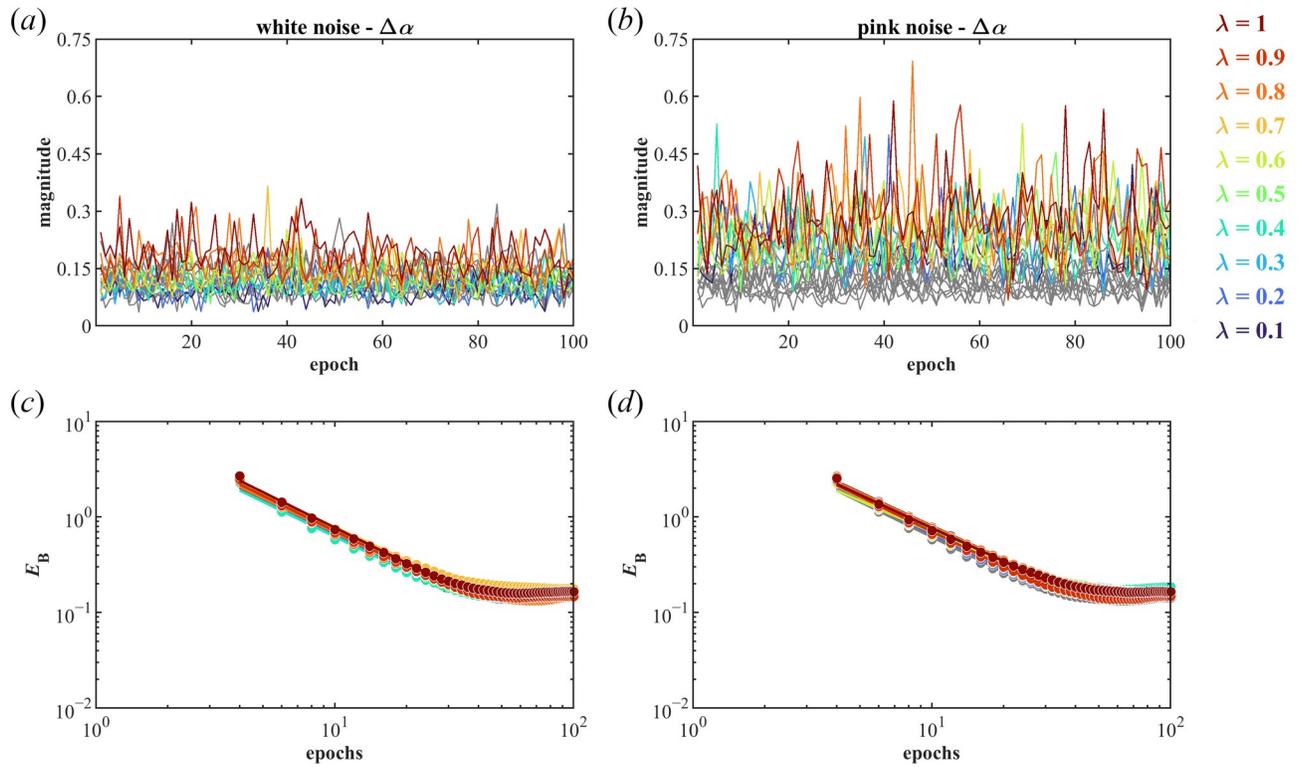

**Figure 8.** Δ$a$ series for the original unsigned pink noise is ergodic irrespective of the level of non-Gaussianity. Ergodicity breaking in Δ$a$ series (Δ$a$ calculated across 500-samples epochs) for unsigned white noise (*a*), and unsigned pink noise (*b*), with different levels of non-Gaussianity. Δ$a$ series for the original unsigned pink noise is higher than the Δ$a$ series for the shuffled unsigned pink noise due to the presence of temporal correlations in the original unsigned pink noise, but no difference exists between Δ$a$ series for the original and shuffled unsigned white noise. Notably, Δ$a$ does not depend on the series' non-Gaussianity. $E_B$ vs. epochs (for lag Δ = 2 epochs) for Δ$a$ series for the original and shuffled unsigned white noise (*c*), and the original and shuffled unsigned pink noise (*d*), with different levels of non-Gaussianity. $E_B$ for Δ$a$ series decreases steadily with epochs and then levels off at larger values at epochs. What emerges is that Δ$a$ series shows none of the ergodicity breaking that the raw series for the original unsigned pink noise does in figure 3. The decay in $E_B$ vs. epochs only differs slightly between Δ$a$ series for the original and shuffled unsigned white noise. It shows almost no difference between Δ$a$ series for the original and shuffled unsigned pink noise. Notably, $E_B$ vs. epochs curves for Δ$a$ series for the original unsigned pink noise does not depend on the series' non-Gaussianity.

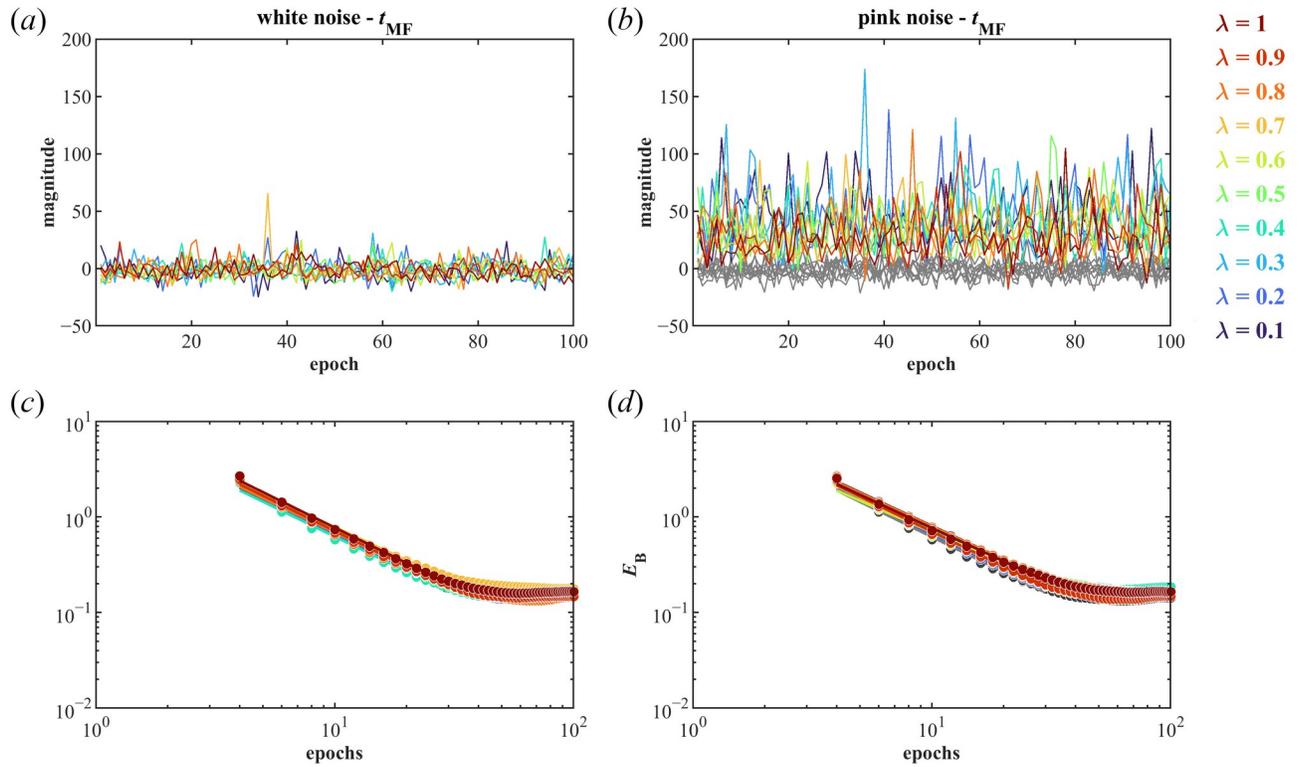

**Figure 9.** $t_{MF}$ series for the original unsigned pink noise is ergodic irrespective of the level of non-Gaussianity. Ergodicity breaking in $t_{MF}$ series ($t_{MF}$ calculated across 500-samples epochs) for unsigned white noise (*a*), and unsigned pink noise (*b*), with different levels of non-Gaussianity. $t_{MF}$ series for the original unsigned pink noise is higher than the $t_{MF}$ series for the shuffled unsigned pink noise due to the presence of temporal correlations in the original unsigned pink noise, but no difference exists between $t_{MF}$ series for the original and shuffled unsigned white noise. Notably, $t_{MF}$ does not depend on the series' non-Gaussianity. $E_B$ vs. epochs (for lag $\Delta = 2$ epochs) for $t_{MF}$ series for the original and shuffled unsigned white noise (*c*), and the original and shuffled unsigned pink noise (*d*), with different levels of non-Gaussianity. $E_B$ for $t_{MF}$ series decreases steadily with epochs and then levels off at larger values at epochs. What emerges is that $t_{MF}$ series shows none of the ergodicity breaking that the raw series for the original unsigned pink noise does in figure 3. The decay in $E_B$ vs. epochs only differs slightly between $t_{MF}$ series for the original and shuffled unsigned white noise. It shows almost no difference between $t_{MF}$ series for the original and shuffled unsigned pink noise. Notably, $E_B$ vs. epochs curves for $t_{MF}$ series for the original unsigned pink noise does not depend on the series' non-Gaussianity.

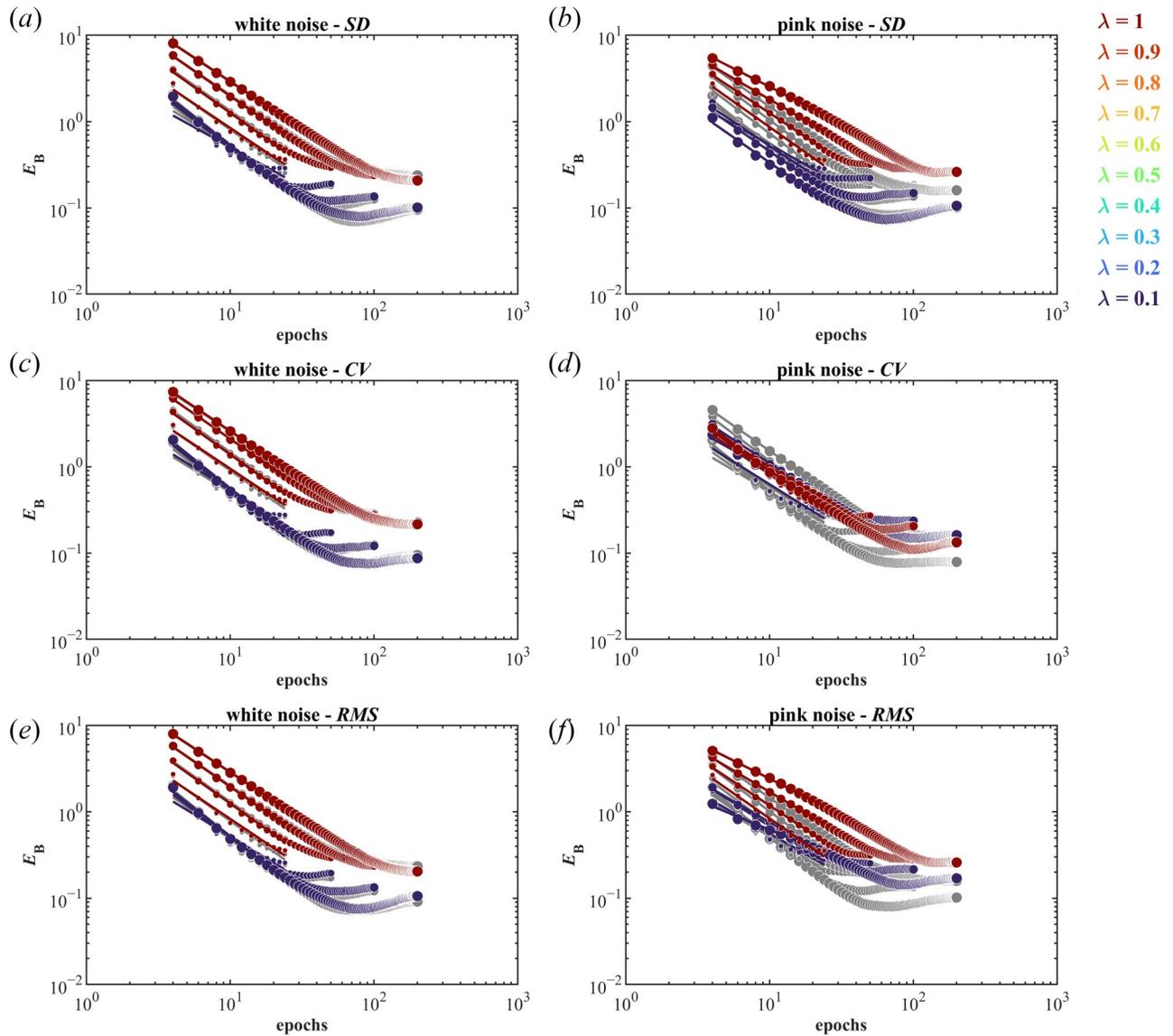

**Figure 10.** $E_B$ vs. epochs curves for epochs of different sizes, showing that the results in figures 4–6 are highly sensitive to epoch size. Each panel shows $E_B$ vs. epochs for 25, 50, 100, and 200 epochs of 2000, 1000, 500, and 250 samples each (small, medium, large, and extra large circles) for noises with very low ($\lambda = 0.1$; dark blue) and very high ($\lambda = 1$; dark red) levels of non-Gaussianity. Noises with intermediate levels of non-Gaussianity ($\lambda = 0.2$–$0.9$) have beeen excluded for clarity. The original unsigned white noise shows no difference from its shuffled counterpart in $E_B$ vs. epochs for (*a*) *SD* series, (*b*) *CV* series, and (*c*) *RMS* series. The $E_B$ vs. epochs curves for white noise does not depend on epoch size for low non-Gaussianity with increasingly depended on epoch size for high non-Gaussianity. In contrast, the original unsigned pink noise show slower decay than its shuffled counterparts in $E_B$ vs. epochs for (*d*) *SD* series, (*e*) *CV* series, and (*f*) *RMS* series. The $E_B$ vs. epochs curves for pink noise depended on epoch size and this dependence was more prominent for high non-Gaussianity. These trends suggest that the *SD*, *CV*, and *RMS* series for white noise break ergodicity only at high non-Gaussianity but the *SD*, *CV*, and *RMS* series for pink noise break ergodicity irrespective of the series' non-Gaussianity. Notably, this ergodicity breaking behavior depends on the chosen epoch size.

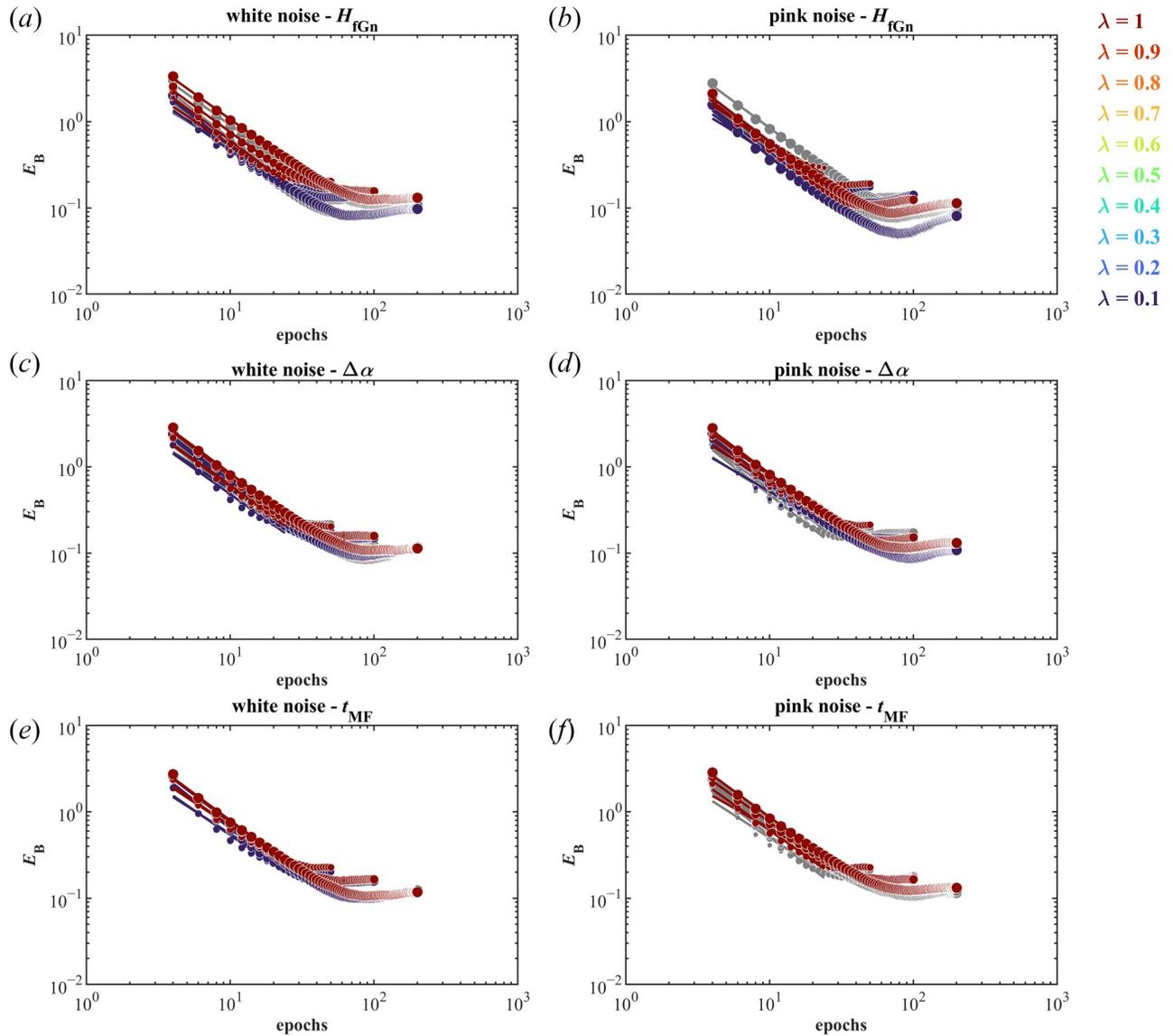

**Figure 11.** $E_B$ vs. epochs curves yielded by the Thirumalai-Mountain method for epochs of different sizes, showing that the results in figures 7–9 are largely insensitive to epoch size. Each panel shows $E_B$ vs. epochs for 25, 50, 100, and 200 epochs of 2000, 1000, 500, and 250 samples each (small, medium, large, and extra large circles) for noises with very low ($\lambda = 0.1$; dark blue) and very high ($\lambda = 1$; dark red) levels of non-Gaussianity. Noises with intermediate levels of non-Gaussianity ($\lambda = 0.2$–$0.9$) have beeen excluded for clarity. The original unsigned white noise shows no difference from its shuffled counterpart in $E_B$ vs. epochs for (a) $H_{fGn}$ series, (b) $\Delta\alpha$ series, and (c) $t_{MF}$ series, for different epoch sizes and non-Gaussianity. The original unsigned pink noise also shows no difference from its shuffled counterpart in $E_B$ vs. epochs for (d) $H_{fGn}$ series, (e) $\Delta\alpha$ series, and (f) $t_{MF}$ series, for different epoch sizes and non-Gaussianity. Minor exceptions are that the $E_B$ vs. epochs curves for $\Delta\alpha$ show greater dependence on epoch size for very high non-Gaussianity in white noise, and the $E_B$ vs. epochs curves are higher for very high non-Gaussianity ($\lambda = 1$; dark red) than very low non-Gaussianity ($\lambda = 0.1$; dark blue) in both white noise and pink noise. These trends suggest that the $H_{fGn}$, $\Delta\alpha$, and $t_{MF}$ series for both white noise and pink noise are ergodic independent of the series' non-Gaussianity and the chosen epoch size.